\documentstyle[sprocl]{article}

\def\be{\begin{equation}}
\def\ee{\end{equation}}
\def\bea{\begin{eqnarray}}
\def\eea{\end{eqnarray}}
\begin{document}

%To Prof Nick Karayiannis -- do read this:-
%If needed the word of Chapter~1, you can type in at the
%\title{}. The words will be in caps and lowercase.
%For chapter title can be in all caps or in caps and lowercase.
%It is up to the author to type for the case sensitive but
%all articles must be in the same style.
%But mostly for Review Volume are without this Chapter~1.
%Thank you
%Jessie   13/4/2000

\title{The role of baryon and lepton numbers in the physics
beyond the standard model
\footnote{Based on the plenary talk given at SUSY'01, Dubna, Russia, and the talk given at Gran Sasso Extended Workshop on Astroparticle Physics, July 2002}}

\author{GORAN SENJANOVI\' C}

\address{International Center for Theoretical Physics,
Trieste, Italy. \\ E-mail: goran@ictp.trieste.it}


\maketitle

\abstracts{I review the role played by baryon and lepton numbers
and their discrete subgroups in determining the low energy
effective theory relevant for TeV physics.}

\section{Introduction}

In a search for the physics beyond the standard model (BSM),
it is useful to use the tool of effective field theory. The
great phenomenological success of the standard model (SM),
based on local gauge invariance and renormalizability, tells us that
the new physics at the scale $M$ should be in the form of
nonrenormalizable operators, cut-off by inverse powers of $M$.
In other words,

\begin{equation}
\label{lagr}
{\cal L}_{\rm effective}={\cal L}(d=4)+\sum_n{1\over M^n}O^{(4+n)}\;,
\end{equation}

\noindent
where $O^{(4+n)}$ denotes operators of mass dimension $4+n$. Two
such operators stand out \cite{weinberg}:

\begin{eqnarray}
\label{oper}
O_p &=&{1\over M_p^2}(q q q l)\;,\nonumber\\
O_\nu &=&{1\over M_\nu}(l l)\Phi^2\;,
\end{eqnarray}

\noindent
where I use a symbolic notation and $q$, $l$ and $\Phi$ stand
for quarks, leptons and the Higgs doublet of the SM. These two
operators have in a sense characterized the two decades of BSM.
They measure violation of $B$ and $L$ symmetries, the accidental
symmetries of the minimal SM. The first one, $O_p$, leads to
proton decay. $(\tau_p)_{exp}>10^{33}$ yrs tells that $M_p$ must
be enormous: $M_p>10^{15}$ GeV. $O_\nu$ leads to neutrino mass
and from $m_\nu\le 1$ eV even $M_\nu$ is large: $M_\nu > 10^{13}$ GeV
(throughout my talk I am assuming only three light neutrinos and
the smallness of $\Delta m^2$ measured by atmospheric and solar
neutrino puzzle, together with the direct $\beta$ decay limit on
$m_{\nu_e}$ implies {\em all} $m_\nu\le 1$ eV).

Thus, even if you ignore gravity, (\ref{lagr}) and (\ref{oper})
indicate immediately the hierarchy problem, the problem of
keeping $M_W/M_p$ and $M_W/M_\nu$ small in perturbation theory
and on a deeper level the problem of understanding the origin of
such small numbers.
 It is normally argued that in any case we have
a hierarchy problem $M_W/M_{Pl}\approx 10^{-17}$, but the Planck
scale $M_{Pl}\approx 10^{19}$ GeV is {\em not} a fundamental scale
(whereas $M_W$ is). This was used by Arkani-Hamed et al. (ADD) in their
proposal \cite{add} that the fundamental scale $M_F$ (the field
theory cut-off, or string scale, or ...) is of the order $1-10$
TeV, and the Planck scale is large due to the large size of extra
space dimensions. In a sense this also poses a (different type of)
hierarchy, but remarkably enough the program is phenomenologically
consistent.

In view of the above it is fair to say that the understanding of proton
stability and the
smallness of the neutrino mass are, in my opinion, the main issues
in the ADD proposal.
A comment is called for regarding the smallness
of neutrino mass. It is often stressed that large extra dimensions
may account naturally for small neutrino Dirac Yukawa coupling
\cite{neutrextra}, {\em i.e.}
small neutrino Dirac mass. This is true, but although this is an
interesting issue, the smallness of Yukawa couplings is NOT a problem
due to protective chiral symmetries.
In a technical sense, small Yukawa couplings
are natural. The real problem is the
smallness of $O_\nu$ is (\ref{oper}) if $M_\nu\approx 1-10$ TeV.
I will come to this later.

If you (as I do), on the other hand, believe in large physical
scale BSM, you can assume a low energy supersymmetry, with the
SUSY breaking scale $\Lambda_{SS}=$ TeV. Amazingly enough, it leads
to the unification of gauge couplings at $M_X\approx 10^{16}$ GeV
\cite{unification}.
It is hard not to be amazed by this fact, for not only the couplings
unify, but they unify in a tiny allowed window. Namely, experiment
and calculability require

\begin{equation}
\label{mxmpl}
10^{15}<M_X<<M_{Pl}\;.
\end{equation}

It is worth stressing that twenty years ago this required predicting
a heavy top quark with $m_t\approx 200$ GeV \cite{ms}.
It was subsequently shown \cite{polchinski}
that a large top Yukawa coupling would cause a flip in the sign
of the Higgs mass when running down from
large ($\approx M_X$ or $M_{Pl}$) to low ($\approx M_W$)
scale. This would explain the great mystery behind the Higgs mechanism.
The price that needs to be paid is the prediction of a desert in a
sense of any new physics all the way from $M_W$ to $M_X$. This
is hard to swallow.

Furthermore, low energy SUSY, just
as large extra dimensions, suffers from the potential catastrophe
regarding proton stability. Again, a natural scale for $M_p$ lies in the
TeV region: $M_p\approx \Lambda_{SS}$. A way out is to assume a so-called
R parity (or matter parity), but this is not so appealing. I will
discuss this at length in what follows and argue that it may be naturally
tied up to the smallness of the neutrino mass.

\section{$B$ and $L$ as probes of BSM}

We see that $B$ and $L$ play clearly a special role in probing
new physics.

Let us focus on low energy SUSY. We need a symmetry which can forbid
some (or all) of $B$ and $L$ violating terms of the supersymmetric
standard model (SSM). Even if you use a discrete symmetry, presumably
such a discrete should eventually be gauged in order to be safe from
gravitational or other high energy physics effects. Iba\~ nez and Ross
\cite{ir}
studied systematically all such anomaly free symmetries, {\em i.e.} all
discrete symmetries that can be embedded in some U(1) (or bigger)
gauge symmetry. They find only two types of such symmetries:

\noindent
(a) any discrete subgroup of U(1)$_{B-L}$ symmetry if you allow
for a right-handed neutrino superfield;

\noindent
(b) a complicated $Z_3$ matter parity. By a suitable U(1)$_Y$
transformation you can rewrite this as a $Z_9$ subgroup of a
baryon number symmetry

\begin{equation}
q\to e^{i{2\pi\over 9}n}q\;.
\end{equation}

Since this symmetry is vector-like, it is free from QCD
anomalies or a $Z^3$ anomaly, and the $Z$ SU(2)$^2$ anomaly is
given by an operator

\begin{equation}
(QQQL)^3\;,
\end{equation}

\noindent
which is obviously invariant under $Z_9$ ($Q$ and $L$ stand for
quark and lepton left-handed doublets separately). Of course,
if you wish to gauge it, you need to gauge $B$ and this requires
new fermions in order to cancel anomalies.

In case (a) a particularly interesting subgroup of U(1)$_{B-L}$
is a $Z_2$ discrete matter parity

\begin{equation}
M=(-1)^{3(B-L)}\;.
\end{equation}

It can be written as $M=R(-1)^{2S}$, where $S$ is spin and $R$
is the usual $R$-parity, under which particles are invariant and
sparticles change sign.

It is, of course, automatic in any theory with $B-L$ gauge symmetry
\cite{rsym}.
The question, the central question in this case, is whether $M$
survives the necessary symmetry breaking down to SSM. Since the gauging
of $B-L$ requires $\nu_R$, the most natural scenario is to give
$\nu_R$ a large mass and decouple it from low-energies. This is
the see-saw mechanism of neutrino mass \cite{seesaw}.

But then, it can be shown that $M$ (or $R$)
remains exact to all orders in perturbation theory \cite{r}.
Thus the usual see-saw mechanism, based on renormalizable field
theory, implies the exactness of $R$-parity.

The proof is straightforward. In MSSM the spontaneous breaking of
$M$ implies the breaking of $L$ and thus the existence of a Majoron
coupled to the $Z$ boson \cite{am}. This is ruled out from the
$Z$ decay width. But the effective field theory argument tells us
that in the limit $M_\nu\to\infty$ (here $M_\nu$ is the scale of
$B-L$ breaking) we have MSSM with the same Majoron. In other words,
there is a pseudo-Majoron with a mass

\begin{equation}
m_J^2\approx {\Lambda_{SS}^3\over M_\nu}<<m_Z^2\;.
\end{equation}

Again, $Z$ decay width rules out such a possibility. Q.E.D.

Once again, just as in the formulas (\ref{lagr})-(\ref{mxmpl})
we have simply used the decoupling theorem. The result is remarkable:
the see-saw mechanism, in which $m_{\nu_R}$ originates in
spontaneous breaking of $B-L$ symmetry at the tree level,
determines the structure of SSM in the form of MSSM. The conservation
of $R$-parity is important for the stability of LSP.

\section{SO(10)}

A good theory should determine $M_\nu$. For this one must go
to Pati-Salam or SO(10). It is worth summarizing here the nice
features of the supersymmetric SO(10) theory with a renormalizable
see-saw mechanism \cite{charan}:

(i) unification of a fermionic family into 16-dim. spinors;

(ii) $M=C^2$, where $C$ is the center of SO(10), {\em i.e.}
$16\stackrel{C}\rightarrow i16$, $10\stackrel{C}\rightarrow
-10$, etc.;

(iii) $R$-parity (or $M$) remain exact even after SU(2)$\times$U(1)
breaking. This is due to the argument described above, {\em i.e.} to
the fact that otherwise there would be a pseudo-Majoron coupled
to the $Z$ boson. The $Z$ decay width rules this out;

(iv) the unification scale $M_X$ is related to the intermediate
scale $M_\nu$. {\em A decrease of $M_\nu$ implies a decrease
of $M_X$};

(v) $M_X>10^{15.5}$ GeV ($\tau_p>10^{33}$ yrs) implies
$M_\nu>10^{13}$ GeV;

(vi) $p\to e^+\pi^0$ is potentially observable and would indicate
the intermediate scale $M_\nu$;

(vii) if a non-canonical see-saw dominates, one can show that
the large mixing angle for atmospheric neutrinos is a consequence
of $b-\tau$ unification \cite{Bajc:2001fe}; and furthermore the leptonic mixing angle is predicted to be $U_{e3}\simeq .16$, to be tested in the near future
\cite{Goh:2003sy}.

(viii) the loss of asymptotic freedom above $M_X$ results in the
gauge coupling becoming strong at the scale $\lambda_S\approx 10M_X$.
This would indicate a string scale below $M_{Pl}$.
For a possible dynamical symmetry brealing scenario, see\cite{aulakh1}.

For a detailed discussion of this theory, see the talk by C. Aulakh
at this conference \cite{aulakhsugra}.

\section{Low cut-off theories}

How to understand the proton stability and the smallness of $m_\nu$
with $M_p\approx M_\nu\approx 1-10$ TeV? In the absence of some
conspiracy, we would need to forbid the dangerous operators of the
type

\begin{equation}
{qqql\over M_p^2}\sum_{n=0}^?c_n\left({M_W\over M_p}\right)^n
\end{equation}

\noindent
for the proton's sake. With $M_p\approx$ TeV and
$c_n\approx O(1)$, it is easy to see that you need to forbid
all the operators up to $n=26$! For $M_p\approx 10$ TeV, you get
$n=12$, still hard to swallow. For neutrino mass, $n=11$ for
$M_\nu\approx$ TeV and $n=5$ for $M_\nu\approx 10$ TeV.

We clearly need a protection mechanism. For example, quarks and leptons
could be living on different branes \cite{famonbran}, ensuring proton
stability, or somehow $B$ and $L$ could be accidental symmetry,
such as in the talk by Iba\~ nez~\cite{ibasusy} at this conference.
  Even more appealing possibility is to have B and L gauged in extra
dimensions as suggested by ADD\cite{add}.
 If the associated gauge bosons live
in the bulk, a remarkable consequence emerges: a strong repulsive force
at distances less than about a millimeter, independently of the number
of extra dimensions 
(forces mediated by the bulk gauge fields may be weakened due to
higher dimensional analog of Meissner effect, provided baryon
number is spontaneously broken on our brane\cite{bulk}).

The authors of Ref \cite{Appelqvist} argue that in 6 dimensions
Lorentz structure and gauge invariance do the job of stabilizing the
proton while the compactification does not spoil it. It is also argued
that the same can be achieved with a $B-L$ symmetry in 6 dimensions;
after compactification one is left with a residual discrete symmetry
that keeps the proton stable while simultaneously ending up with a
small neutrino mass \cite{Mohapatra-PerezL}

It is worth, thus, to systematically study the issue from the
field theory point of view. What is the minimal prize needed to
gauge $B$ and $L$ and ensure the absence of (\ref{oper})?
Surprisingly enough, there is still no systematic study.

The simplest possibility is to add three more, but now anti,
families, {\em i.e.} three families with $V+A$ couplings (mirrors)
\footnote{See also the talk by Kobakhidze~\cite{archil}.}.
This is probably ruled out phenomenologically from the high
precision electro-weak data, and it also require a mirror
(gauge) symmetry needed to prevent the pairing off of ordinary
and mirror fermions.

You could try adding a single mirror-like family, but with different
$B$ and $L$ charges. The problem are $B^3$ and $L^3$ anomalies.
I shall not bore you with the details here; suffice it to say
that it seems very artificial and hard to gauge $B$ and $L$ in
4 dimensions.

\section{Summary and outlook}

We have seen that baryon and lepton numbers play an important
role in the physics beyond the standard model. In a sense, they
could give us a clue of what lies ahead. Supersymmetric grand unification
illustrates perfectly this point. Here the unification of gauge
couplings is a true physical phenomenon implying the single gauge
coupling above $M_X\approx 10^{16}$ GeV. At the
same time $M_X$ is tied with the proton lifetime: $\tau_p
(p\to \pi^0 e^+)\approx M_X^4$, with $\tau_p\approx 10^{35\pm 1}$
yrs.

Furthermore, a lepton number violation at the scale $M_\nu\le
M_X$ gives a small mass for neutrinos and can imply the exactness
of $R$ parity and LSP as a dark matter candidate.

On the other hand d=5 operators tend to imply, through
$\tau_p\propto M_X^2$, a rather short proton lifetime $\tau_p\le
10^{32}$ yrs (see for example \cite{murayama}). This,
however is model dependent, {\em i.e.} it depends on the form of
fermion (see for example \cite{zurab}) and sfermion
masses matrices, and on the uncertainties in the unification scale. 
It is therefore premature to
claim a death blow even for a minimal SU(5)
theory \cite{death}. For a careful analysis of this
issue see \cite{life}.

There is a generic trouble, though,
through non-renormalizable d=5 operators. If you write
such a term in the superpotential, cut-off by the Planck
scale

\begin{equation}
\label{superp}
{\cal W}_{\rm effective}={1\over M_p}(QQQL);,
\end{equation}

\noindent
where $Q$ and $L$ denote the quark and lepton superfields, respectively,
the proton lifetime turns out to be too fast by some 12 or more orders
of magnitude. Is this an argument in favor of gauging B and L? In this
case I would be willing, although reluctantly, to buy the possibility
of the 1 - 10 TeV field theory cut-off a la ADD.

\section*{Acknowledgments}

I thank the organizers of SUSY'01 for a warm hospitality and an
excellent conference. I am grateful to my collaborators Charan
Aulakh, Borut Bajc, Alejandra Melfo, Pavel Fileviez-P\'erez, Andrija Ra\v sin 
and Francesco Vissani. I also wish to acknowledge useful discussions
with Gia Dvali.
This work is partially supported by EEC
under the TMR contracts ERBFMRX-CT960090 and HPRN-CT-2000-00152.

\end{document}